\documentclass[letterpaper,10pt]{article} 

\usepackage{osameet3}

\usepackage{amsmath,amssymb}
\usepackage[colorlinks=true,bookmarks=false,citecolor=blue,urlcolor=blue]{hyperref} 
\usepackage{graphicx,epstopdf}

\begin{document}

\title{Experimental demonstration of arbitrary Raman gain--profile designs using machine learning}

\author{Uiara~C.~de~Moura$^1$, Francesco~Da~Ros$^1$, A.~Margareth~Rosa~Brusin$^2$, Andrea~Carena$^2$, and Darko~Zibar$^1$}
\address{1. DTU Fotonik, Technical University of Denmark, DK-2800, Kgs. Lyngby, Denmark\\2. DET, Politecnico di Torino, Corso Duca degli Abruzzi, 24 - 10129, Torino, Italy}
\email{uiamo@fotonik.dtu.dk}

\copyrightyear{2020}

\begin{abstract}
A machine learning framework for Raman amplifier design is experimentally tested. Performance in terms of maximum error over the gain profile is investigated for various fiber types and lengths, demonstrating highly--accurate designs.

\end{abstract}
\ocis{(060.0060) Fiber optics and optical communication, (060.2320) Fiber optics amplifiers}

\section{Introduction}
To satisfy future data capacity demands, next generation of optical communication systems is envisioned to operate in O+E+S+C+L band~\cite{Napoli18}. One of the challenges with ultra--wideband transmission is the design of optical amplification schemes. Currently, there are several proof--of--principle solutions relying on bismuth--doped fiber amplifiers (BDFA), semiconductor optical amplifier (SOA) and hybrid--Raman amplifiers, being investigated by various groups~\cite{Manuylovich19, Renaudier17, Iqbal19, Ionescu19}.

The advantage of Raman amplifiers is that they can provide flexible gain profiles in O, E, S, C and L bands by properly adjusting the pump powers and wavelengths. This is an important feature as Raman amplifiers can then be used in hybrid configurations with erbium-doped fiber amplifiers (EDFA), BDFAs and SOAs to provide gain flattening and in general for gain profile shaping. Moreover, Raman amplifiers can be used in discrete or distributed stand--alone configurations to provide ultra-wideband gain~\cite{Iqbal19, Iqbal18}. The suitable selection of pump powers and wavelengths is the key in achieving a desired Raman gain profile. This is a challenging task as the relationship between gain profile versus pump powers and wavelengths is nonlinear and requires extensive numerical simulations to predict, especially for a large number of pumps~\cite{Neto}.

Recently, a machine learning (ML)--based framework for Raman amplifier design has been proposed~\cite{Zibar19,ZibarJLTsubmitted}. The advantage of the proposed framework in~\cite{Zibar19,ZibarJLTsubmitted} is that it can provide an ultra--fast and low--complexity pump powers and wavelengths allocation for the design of arbitrary Raman gain profiles.

In this paper, extensive experimental investigations of the framework proposed by~\cite{Zibar19,ZibarJLTsubmitted} are presented. We test different fiber types ranging from standard single mode fiber (SSMF), ultra--low loss fiber (ULLF), dispersion compensating fiber (DCF), highly nonlinear fiber (HNLF) and inverse dispersion fiber (IDF), each with a different fiber length, covering distributed and discrete Raman amplifier applications. The accuracy of the ML--based framework in predicting the pump powers to provide a set of arbitrary gain profiles is investigated by comparing the target gain profiles with the profiles experimentally obtained when applying the pump powers predicted by the framework. A vast statistically analysis of several thousand cases is conducted and error distributions are estimated showing that the ML-based framework proposed in~\cite{Zibar19,ZibarJLTsubmitted} can effectively design Raman amplifier with limited errors.

Even though experimental results are shown only for the C--band in this paper, because of limitation in the available Raman pump wavelengths, we believe that main conclusions can be extended to larger bandwidth, i.e. C+L, as we have shown by simulation in~\cite{ZibarJLTsubmitted}. 

\section{Experimental setup}
\begin{figure}[t]
  \centering
  \includegraphics[width=0.9\textwidth]{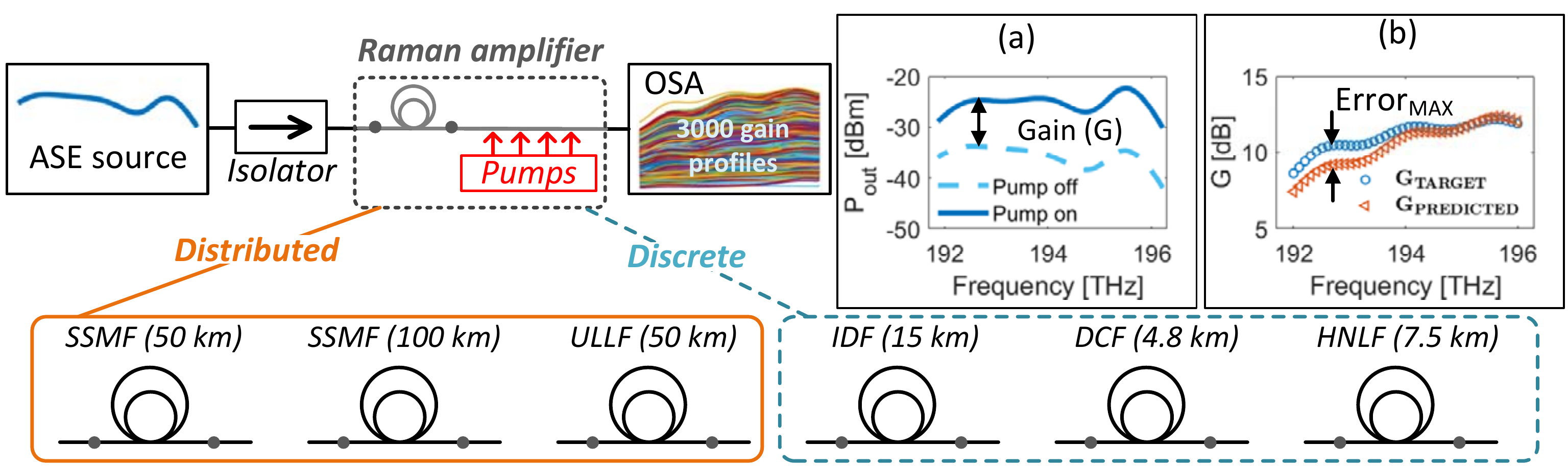}
\caption{Experimental setup to capture the data--set for the NN training and further validations showing the ASE (input) and output spectra, and illustrating how to obtain the (a) the gain and (b) the maximum error ($Error_{MAX}$) from the gain profiles.}
\label{fig:setup}
\end{figure}

Fig.\ref{fig:setup} shows the experimental setup used to generate the data--set for the ML--based Raman amplifier design. The input signal is emulated by an amplified spontaneous emission (ASE) source covering the entire C--band (192-196~THz). Its profile is also depicted in Fig.\ref{fig:setup}. The total ASE power launched into the optical fiber is +2.6~dBm.  An optical isolator is used to prevent the counter--propagating Raman pump power from entering the ASE source.

Five fibers with different Raman gain coefficients are evaluated: 50 and 100~km of SSMF, 50~km of ULLF, 4.8~km of DCF, 7.5~km of HNLF, and 15~km of IDF. SSMF is widely deployed in terrestrial systems while ULLF is used for submarine and unrepeatered links due to its lower loss and wider effective area (although the reduced Raman gain). Both SSMF and ULLF are used as the gain medium for distributed Raman amplifiers.

The DCF and HNLF~\cite{Iqbal19JLT} are special highly nonlinear fibers used for discrete Raman amplifiers, the former presenting the advantage to also compensate for chromatic dispersion. The IDF is being used for discrete Raman amplifiers as an alternative to DCFs due to its lower attenuation~\cite{Iqbal19JLT}. However, its original application is for distributed Raman amplifiers in submarine systems.

The commercial Raman pump module consists of four pumps with adjustable powers and fixed wavelengths. The fixed wavelengths are 1454.4~nm, 1444.8~nm, 1434.4~nm, and 1423.4~nm (being able to amplify the full C--band). The powers can be adjusted to values up to 145~mW, 158.5~mW, 180~mW, and 152.5~mW, respectively. Although we have lost some degrees of freedom to design the Raman amplifier due to the fixed wavelengths, this is a realistic scenario using commercially deployable equipment. At the span end, an optical spectrum analyzer (OSA) captures the spectra that will allow to determine the gain profile.\vspace{0.1cm}

The experimental data--set is generated in the following manner: for each fiber type, values for the Raman pump powers are drawn from an uniform distribution and the corresponding Raman output power spectra are recorded by the OSA (see illustration in Fig.\ref{fig:setup}). The size of the experimental data--set is 3000. The Raman gain is calculated by subtracting the output power spectra ($P_{out}(f)$ [dBm]) with the pumps turned off from the spectra with the pumps turned on, as illustrated in Fig.\ref{fig:setup}(a). This procedure gives the on-off gain of the Raman amplifier which is then discretized over the 40 C--band channels (100-GHz, ITU-T grid).
The data--set is the split in two halves: one for training and the other for final validation.\vspace{0.1cm}

Then, the ML-based framework for pump powers allocations is applied offline. During the training stage, a multi-layer neural network, $NN_{bw}(\cdot)$, is employed to learn the (inverse) mapping between the gain profiles, $\mathbf{G}$ and pump powers $\mathbf{P}$. Thereafter, if the error is not acceptable a fine optimization based on gradient descent is applied to fine--adjust the pump powers and reduce the test error. The gradient descent uses another multi-layer neural network, $NN_{fw}(\cdot)$, representing the (forward) mapping between $\mathbf{P}$ and $\mathbf{G}$~\cite{ZibarJLTsubmitted}.
\vspace{0.1cm}

For $NN_{bw}(\cdot)$, random projection method is applied to learn the weights. The network topology consists of 2 hidden layers, 600 neural nodes per layer and hyperbolic tangent as the activation function. A model averaging using 10 parallel neural networks is used. For $NN_{fw}(\cdot)$ Levenberg-Marquardt algorithm is used to learn the weights. The topology consists of 2 hidden layers with 10 neural nodes, and hyperbolic tangent as the activation function. These topologies for $NN_{bw}(\cdot)$ and $NN_{fw}(\cdot)$ were used for all considered fiber types but NN must be specifically trained for each of them.  

\section{Results and discussions}
The accuracy of the proposed ML-based framework to design an arbitrary Raman gain profile is evaluated in the following way: the desired (target) gain profiles ($\mathbf{G_{TARGET}}$) from the final validation data--set are applied to the framework to obtain the corresponding set of allocated pump powers. These pump powers are then applied to the experimental setup to measure the corresponding (predicted) gain profile ($\mathbf{G_{PREDICTED}}$). The maximum error, $Error_{MAX}$, is defined as the largest deviation between $\mathbf{G_{TARGET}}$ and $\mathbf{G_{PREDICTED}}$ over the whole considered C-band, as illustrated in Fig.\ref{fig:setup}(b). \vspace{0.1cm}

\begin{figure}[t]
  \centering
  \includegraphics[width=1.00\textwidth]{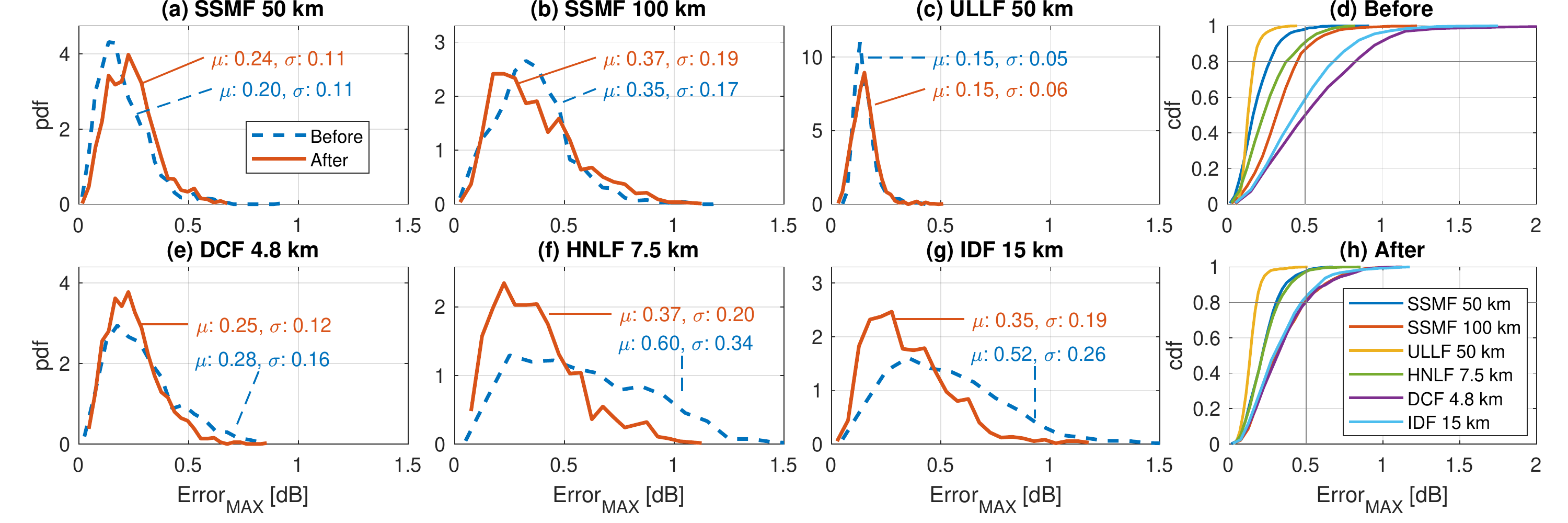}
\caption{Probability density function (pdf) and cumulative distribution function (cdf) of the maximum error ($Error_{MAX}$) between target and designed gains over the validation data--set.}
\label{fig:result}
\end{figure}

Fig.\ref{fig:result} shows the probability density function (pdf) and the cumulative distribution function (cdf) for $Error_{MAX}$ over the final validation data--set for all evaluated fiber types. Two curves have been plotted: before and after fine--optimization. ``Before'', indicates that only inverse ML-NN, $NN_{bw}(\cdot)$, is employed while ``after'' indicates that fine--optimization, relying on gradient descent and $NN_{fw}(\cdot)$, is employed. 

The results in Fig.\ref{fig:result} demonstrate that the $NN_{bw}(\cdot)$ is able to provide accurate pump power allocations for SSMF, ULLF and DCF (Fig.\ref{fig:result}(a-c,e), dashed curves). Mean values for the $Error_{MAX}$ are 0.20 (SSMF-50~km), 0.35 (SSMF-100~km), 0.15 (ULLF), and 0.28~dB (DCF). For these fibers, the cdf curves shows $Error_{MAX}<$ 0.5~dB for 98\% (SSMF 50~km), 87\% (SSMF 100~km), 100\% (ULLF), and 91\% (DCF) of the cases. Applying the  fine--optimization (solid curves) does not introduce significant changes.

On the other hand, using only $NN_{bw}(\cdot)$ does not provides enough accuracy for HNLF and IDF (Fig.\ref{fig:result}(f,g)), where just 50\% and 59\% of the cases have $Error_{MAX}$ $<$ 0.5~dB, according to their cdf curves. In these cases there is a performance improvement after the fine--optimization, with a mean $Error_{MAX}$ reduction from 0.60 to 0.37~dB (HNLF) and from 0.52 to 0.35~dB (IDF). The cdf curves also illustrate the fine--optimization improvements, increasing to 80\% (HNLF) and 83\% (IDF) the cases with $Error_{MAX}<$ 0.5~dB. \vspace{0.1cm}

Overall, we can conclude that ultra--fast, low--complexity and high--accuracy pump power allocation is feasible with only inverse ML-NN, $NN_{bw}(\cdot)$. For those cases where $NN_{bw}(\cdot)$ is not so accurate due to potentially non-optimized topology model, the fine--optimization plays an important role, reducing the maximum error.

\section{Conclusion}
This work reports for the first time the experimental demonstration of pump power allocation for the design of Raman amplifier arbitrary gain profiles employing various fiber types (SSMF, ULLF, DCF, HNLF and IDF). The obtained results show that the proposed machine learning-based framework is able to predict the pump powers with fixed wavelengths to achieve arbitrary gain profiles with a maximum error of 0.5~dB for 80\% of the cases.
\\

\noindent
\textit{Acknowledgements} \\
\noindent
\footnotesize{
This project has received funding from the European Union’s Horizon 2020 research and innovation programme under the Marie Sk\l{}odowska-Curie grant agreement No 754462 and the European Research Council through the ERC-CoG FRECOM project (grant agreement no. 771878).
}

\end{document}